\documentclass{PoS}

\usepackage{amssymb}

\usepackage{multirow}

\usepackage{enumitem}

\usepackage{ulem}

\usepackage{rotating}

\usepackage{natbib}
\setcitestyle{numbers,open={[},close={]}}

\usepackage{latexsym}

\usepackage[hyphens]{url}

\usepackage{array}
\newcolumntype{$}{>{\global\let\currentrowstyle\relax}}
\newcolumntype{^}{>{\currentrowstyle}}
\newcommand{\rowstyle}[1]{\gdef\currentrowstyle{#1}%
  #1\ignorespaces
}

\title{Super-Massive Neutron Stars and Compact Binary Millisecond Pulsars}

\ShortTitle{Super-Massive Neutron Stars}

\author{\speaker{Manuel Linares}$^{ab}$
  \\
  \llap{$^a$}Departament de F{\'i}sica, EEBE, Universitat Polit{\`e}cnica de Catalunya, Av. Eduard Maristany 16, E-08019 Barcelona, Spain.\\
  \llap{$^b$}Institute of Space Studies of Catalonia (IEEC), E-08034 Barcelona, Spain.\\
        E-mail: \email{manuel.linares@upc.edu}}

\abstract{

The maximum mass of a neutron star has important implications across
multiple research fields, including astrophysics, nuclear physics and
gravitational wave astronomy. Compact binary millisecond pulsars (with
orbital periods shorter than about a day) are a rapidly-growing pulsar
population, and provide a good opportunity to search for the most
massive neutron stars. Applying a new method to measure the velocity
of both sides of the companion star, we previously found that the
compact binary millisecond pulsar PSR~J2215+5135 hosts one of the most
massive neutron stars known to date, with a mass of
2.27$\pm$0.16~M$_\odot$ (Linares, Shahbaz \& Casares, 2018). We
reexamine the properties of the 0.33~M$_\odot$ companion star, heated
by the pulsar, and argue that irradiation in this ``redback'' binary
is extreme yet stable, symmetric and not necessarily produced by an
extended source. We also review the neutron star mass distribution in
light of this and more recent discoveries. We compile a list of all
(nine) systems with published evidence for super-massive neutron
stars, with masses above 2~M$_\odot$. We find that four of them are
compact binary millisecond pulsars (one black widow, two redbacks and
one redback candidate). This shows that compact binary millisecond
pulsars are key to constraining the maximum mass of a neutron star.  }

\FullConference{Multifrequency Behaviour of High Energy Cosmic Sources - XIII - MULTIF2019\\
		3-8 June 2019\\
		Palermo, Italy}

\begin{document}

\def \aj {AJ}
\def \mnras {MNRAS}
\def \apj {ApJ}
\def \apjs {ApJS}
\def \apjl {ApJL}
\def \aap {A\&A}
\def \aapr {A\&ARv}
\def \nat {Nature}
\def \araa {ARAA}
\def \pasp {PASP}
\def \aaps {AAPS}
\def \prd {PhRvD}
\def \apss {ApSS}
\def \physrep {PhysRep}

\section{Introduction: the maximum neutron star mass}
\label{sec:intro}


The maximum mass that a neutron star (NS) can sustain, M$_{NS}^{max}$,
has important implications across multiple research fields, including
astrophysics, nuclear physics and gravitational wave (GW) astronomy.
From a purely observational viewpoint, the distinction between neutron
stars and stellar-mass black holes in binary systems is often based on
the mass of the primary star, which is classified as a black hole if
its mass exceeds M$_{NS}^{max}$ \citep{Casares94}.
The birth masses of NSs depend on the still uncertain physical
mechanisms taking place in supernovae, when the proto-NS is being
formed during the gravitational collapse of a massive star
\citep{Timmes96}.
Later evolution in an interacting binary can significantly increase
the mass of a NS via accretion, as well as decrease its spin period
and surface magnetic field, during the so-called recycling stage
\citep{Alpar82}.
More recently, the maximum NS mass has regained interest in the
context of double neutron star mergers and the associated GW emission
\citep{GW170817}.
Simply put, what is left after two neutron stars merge depends on
their masses before coalescing, and on the maximum mass that a neutron
star can hold \citep[e.g.,][]{Margalit17}.
It is also important to know M$_{NS}^{max}$ well enough to distinguish
between {\it massive} NSs and {\it light} black holes, in the new era
of GW events from binary mergers.

The maximum mass of a NS depends on the compressibility of its core,
and on the equation of state (EoS, the relation between pressure and
density) at densities close to and above nuclear saturation density
($\rho_{nuc} \simeq$3$\times$10$^{14}$~g~cm$^{-3}$).
Relatively incompressible matter (described by
``stiff'' EoSs) is needed to support more massive NSs.
Microscopically, the EoS in the core and M$_{NS}^{max}$ are set by the
nucleon-nucleon interactions in the central parts of a NS, which
remain uncertain despite six decades of research \citep{Salpeter60}.
Therefore, the maximum mass of a NS has a critical impact on nuclear
physics, as it places direct constraints on the EoS of matter at
supra-nuclear densities \citep{Page06,Lattimer07,Ozel16}.

The NS mass distribution has evolved as new types of NSs in binaries
were discovered and their masses measured.
The majority of NS mass measurements come from rotation-powered
radio pulsars in binary systems (299 listed at the ATNF
catalog, v1.60; \citep{ATNFpsrcat}), most of which are binary {\it
  millisecond} pulsars (MSPs, 249 catalogued with spin periods shorter
than 30~ms; see also \citep{LorimerCat}).
Two decades ago, the first measurements suggested a
single and narrow NS mass distribution near 1.4~M$_\odot$
\citep[1.35$\pm$0.04~M$_\odot$ from 26 pulsars, mostly in double
  NSs;][]{Thorsett99}.
In 2010 and 2013, NSs with masses close to 2~M$_\odot$ were found in
binary pulsars with white dwarf (WD) companions
\citep{Demorest10,Antoniadis13}.
These were accurate measurements ($\simeq$2\% 1-$\sigma$ errors; see
Table~\ref{table:masses}) obtained with different methods, and they
established that NSs can be as massive as 2~M$_\odot$.
In 2011, van Kerkwijk et al. \citep{Kerkwijk11} found evidence for a
NS with more than two Solar masses, 2.40$\pm$0.12~M$_\odot$, in
%
one of only four compact binary MSPs known in the Galactic field at
the time (see Section~\ref{sec:aspire} for definition and details).

Recently, {\"O}zel \& Freire \citep{Ozel16} compiled NS mass (and
radius) measurements, discussing in depth the NS mass distribution.
Thanks to these and other previous studies \citep{Page06,Lattimer07,Ozel12},
we know now that NS masses span a wide range, between 1.2 and at least
2~M$_\odot$.
However, compact binary MSPs (black-widows and redbacks) have not been
included in most global studies of the NS mass distribution, perhaps
due to their past scarcity or in order to avoid the uncertainties in
correcting for the strong irradiation of the companion star by the
pulsar wind.
Here, we review the most massive NSs presently known in light of
recent findings and including compact binary MSPs, which we argue are
key to constraining M$_{NS}^{max}$
\citep{Kerkwijk11,Romani12b,Breton13,Linares18b,Strader19}.

\begin{figure}[t]
  \begin{center}
\resizebox{1.0\columnwidth}{!}{\rotatebox{-90}{\includegraphics[]{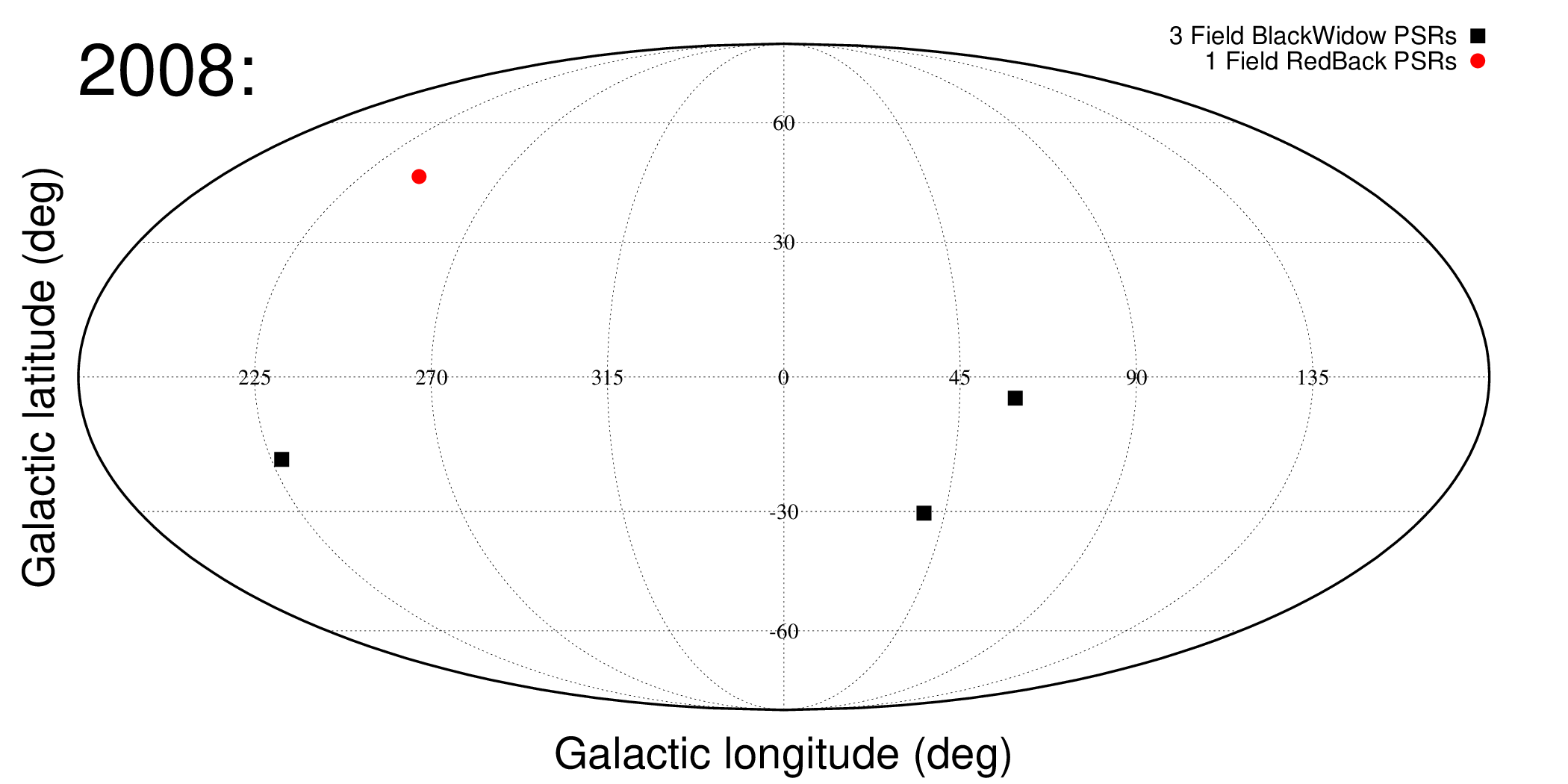}}}
\resizebox{1.0\columnwidth}{!}{\rotatebox{-90}{\includegraphics[]{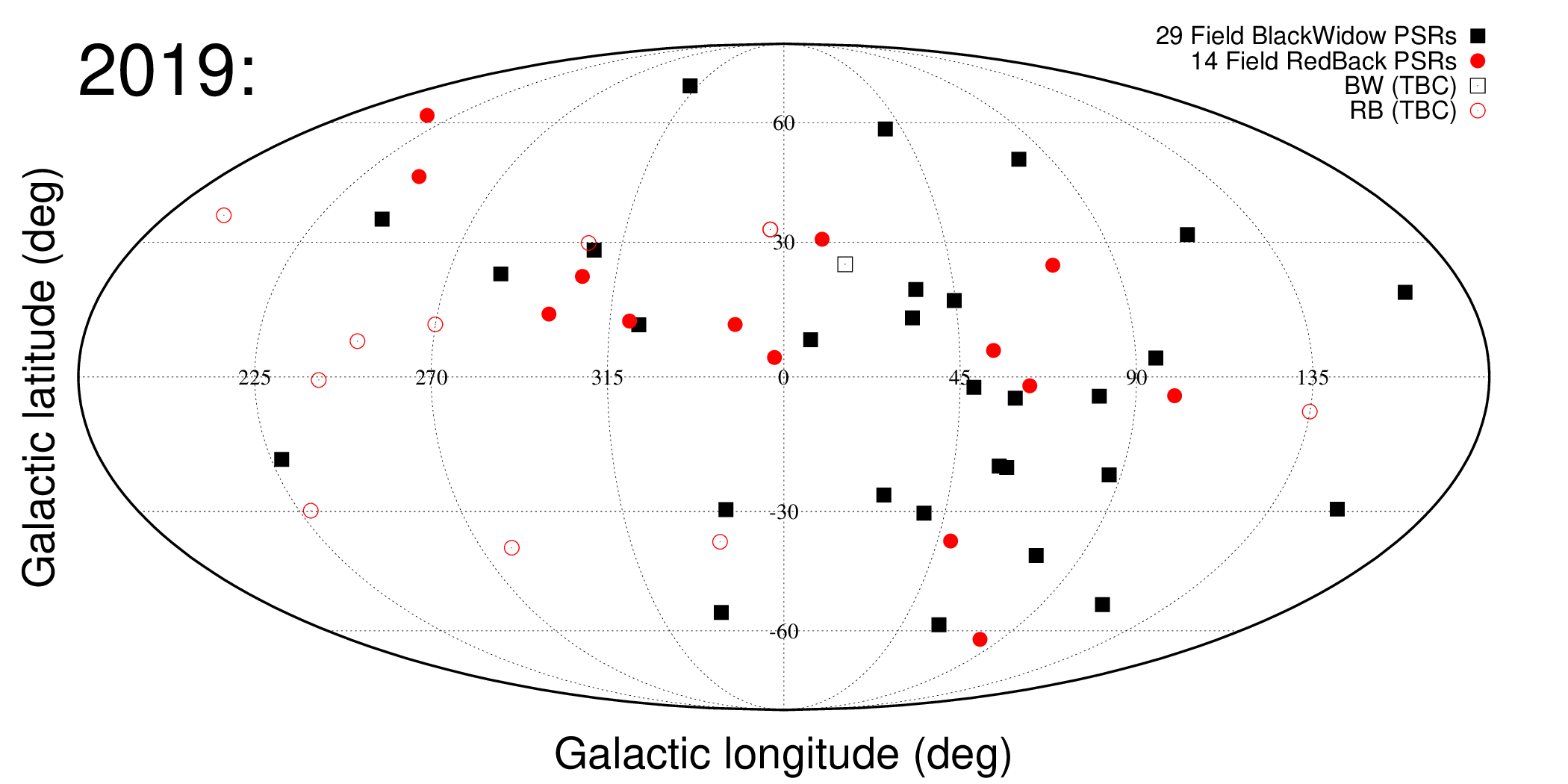}}}
\caption{All-sky map (Mollweide projection) showing the Galactic
  coordinates of the compact binary MSPs known in 2008 (top, 4
  systems) and 2019 (bottom, 43 confirmed + 11 candidates). Black
  squares and red circles show black-widow and redback MSPs,
  respectively. Open symbols show candidate systems, where the
  detection of a radio/gamma-ray pulsar has not been reported yet.}
\label{fig:maps}
\end{center}
\end{figure}

\section{A spider revolution: millisecond pulsars in compact binaries}
\label{sec:aspire}

Compact binary MSPs have orbital periods shorter than about a day and
form two distinct populations, with either a low-mass main-sequence
companion star in the systems known as ``redbacks'' (RBs;
M$_c$$\gtrsim$0.1~M$_\odot$), or a very low-mass ``ultra-light''
companion in those known as ``black widows'' (BWs; M$_c
\sim$~0.01~M$_\odot$).
They have been nicknamed after cannibalistic spiders because of the
destructive effect that the pulsar wind may have on its companion, and
are sometimes referred collectively as ``spiders''.
Before the launch of the {\it Fermi} gamma-ray telescope in 2008, only
four compact binary MSPs were known in the field of the Galaxy, as
shown in Figure~\ref{fig:maps} (top).
These exclude MSPs in globular clusters, which now contain 12/18
RBs/BWs \citep{FreireCat} but are typically unsuited for dynamical
mass measurements.
Pulsars are relatively bright sources at GeV energies, the band where
{\it Fermi}'s large area telescope (LAT, $\sim$0.1--100~GeV;
\citep{Acero15,4FGL}) is providing unprecedented sensitivity.
Combined with targeted radio observing campaigns
\citep{Hessels11,Ransom11,Bates11,Ray12,Stovall14,Camilo15,Bates15,Sanpa16}
as well as optical and multi-wavelength follow-up
\citep{Romani11,Kong12,Romani12,Romani14,Li16,Linares17}, this has
turned {\it Fermi}-LAT into a true ``pulsar discovery machine''.

Surprisingly, many of these LAT-driven discoveries have revealed new
compact binary MSPs, so their population has increased drastically
over the past decade, from 4 to more than 40 (Figure~\ref{fig:maps},
bottom).
This is what we may call ``a spider revolution'': a new class of nearby
(d$\lesssim$3~kpc) and energetic (spin-down luminosity $\dot{E}
\sim$10$^{34}$--10$^{35}$~erg~s$^{-1}$) pulsars has emerged.
Because most new spiders are also far from the Galactic plane, where
interstellar extinction is low, they are well within reach of optical
telescopes (especially redbacks, with brighter companion stars).
This allows for spectroscopic studies of the companion star and
dynamical measurements of the MSP mass, as detailed below.
%
During a long (Gyr) formation phase with active mass transfer, spiders
are thought to accrete mass from their companion stars, so that NSs in
compact binary MSPs may be significantly more massive than they were
at birth \citep{Chen13}.
Thus, among other fields, this new MSP population is having a strong
impact on M$_{NS}^{max}$ and the NS mass distribution.

\section{A 2.3 Solar-mass neutron star in PSR~J2215+5135}
\label{sec:j2215}

In 2014, we set out to measure the mass of a redback pulsar in a
4.14~hr orbit, PSR~J2215+5135, which had previous evidence for a
massive NS and a strongly irradiated companion star
\citep{Breton13,Schroeder14}.
We observed the system with three optical telescopes, obtaining high
signal-to-noise photometry and including the Gran Telescopio Canarias
(GTC, 10.4-m diameter) in order to collect high-quality spectra
throughout the orbit.
Thanks to these GTC spectra we were able to find a new and subtle
effect: among the absorption lines from the companion star's
atmosphere, metallic lines trace the unheated ``cold'' side of the
star and thus move at higher velocities than hydrogen lines, which are
formed predominantly on the inner ``hot'' side of the star.
In particular, we found that magnesium lines (MgI triplet at
5167-5184~\AA) move 10\% faster than Balmer lines
(H$\beta,\gamma,\delta$).
This allowed us to find empirically and robustly the velocity of the
center of mass of the star (K$_2$), a very important step to obtain
reliable mass measurements in irradiated binary systems.

Furthermore, we measured the temperature of the star throughout the
orbit using a suite of absorption lines and confirmed quantitatively
the extreme irradiation of the inner side of the companion produced by
the pulsar wind (from T$_\mathrm{N}$=5660$^{+260}_{-380}$~K to
T$_\mathrm{D}$=8080$^{+470}_{-280}$~K).
Imposing these independent temperature constraints, we modeled jointly
the light and radial velocity curves in order to obtain the orbital
inclination (i=63.9$^\circ\pm$2.5) and the masses of both stars.
While our work was being completed, Romani et al. reported higher
inclinations from independent dynamical studies of PSR~J2215+5135
\citep{Romani15,Romani16}.
Because their reported temperatures differ from ours and have no
uncertainties, we concluded that accurate temperature constraints are
critical to measure robustly masses and inclination.
For the full details and discussion the reader is referred to Linares,
Shahbaz \& Casares (2018, \citep{Linares18b}).

\begin{figure}[h!]
  \begin{center}
\resizebox{1.0\columnwidth}{!}{\rotatebox{-90}{\includegraphics[]{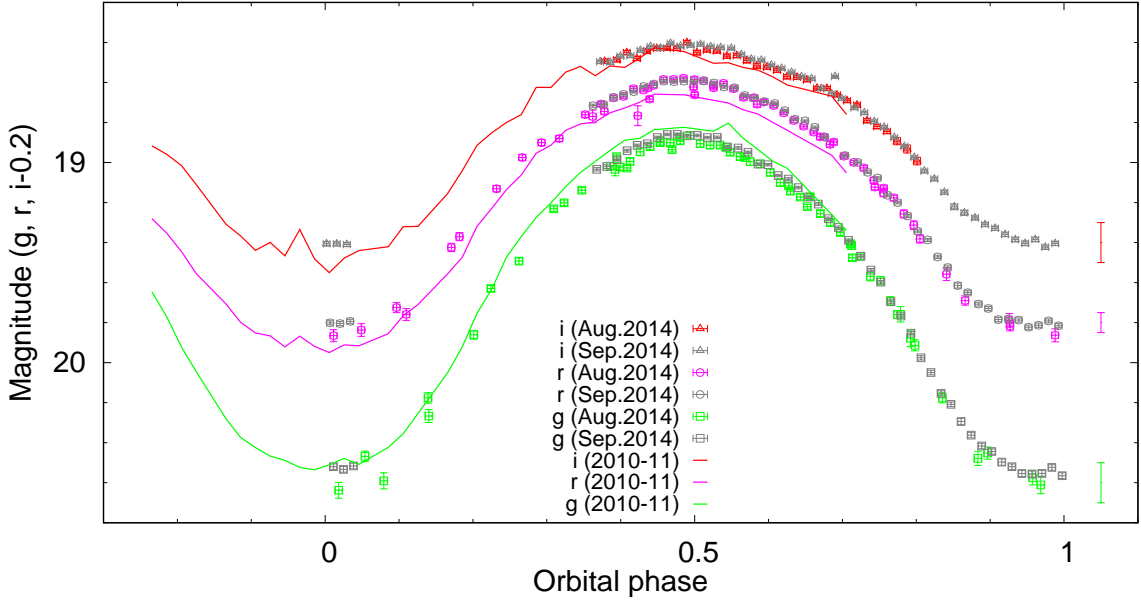}}}
\caption{Optical light curves of PSR~J2215+5135 taken during four
  different nights with the WHT and IAC80 telescopes on August (red
  triangles, magenta circles and green squares for the i, r and g
  bands, respectively) and September (gray symbols), 2014. Solid lines
  show the light curves measured by Schroeder \& Halpern in 2011 (3
  years earlier, converted from BVR filters; \citep{Schroeder14}). The
  orbital light curves of this redback MSP are smooth (down to the
  shortest 60~s exposures) and stable on timescales of days to years
  (within the available data and its photometric precision, shown by
  the error bars on the right).}
\label{fig:j2215}
\end{center}
\end{figure}

We thereby found one of the most massive NSs known to date in PSR
J2215+5135, with M$_\mathrm{NS}$=2.27$\pm$0.16~M$_\odot$
\citep{Linares18b}.
A 2.3 Solar-mass neutron star rules out most currently proposed
equations of state, casting doubt on the existence of exotic forms of
matter in the core.
It is worth stressing here that the optical light curves and spectra
of PSR~J2215+5135 are ``well behaved'', in the sense that they only
show the orbital modulation typical of irradiated compact binaries. No
flares, long-term variability or transient emission lines have been
detected (see, e.g., multi-epoch light curves in
Figure~\ref{fig:j2215}).
We were able to fit jointly and satisfactorily these three-band light
curves (g,r,i) as well as the two-species radial velocity curves
(Balmer and MgI) assuming point-like irradiation by the pulsar wind,
without any extended heating source (unlike previous work
\citep{Romani16}).
Furtermore, after carefully considering the uncertainty in the orbital
phases, we did not find in our data significant asymmetry in the
optical lightcurves of PSR~J2215+5135 \citep{Schroeder14,Romani15}.
Summarizing, in our optical study of this particular redback we found
that even if heating/irradiation of the companion by the MSP is
extreme: i) it is not variable, ii) it is not necessarily produced by
an extended region and iii) it shows little or no asymmetry.
We argue that this, together with our new empirical method to
determine K$_2$, makes the 2.3~M$_\odot$ measurement robust.

\begin{figure}[p]
  \begin{center}
    \includegraphics[height=0.95\textheight]{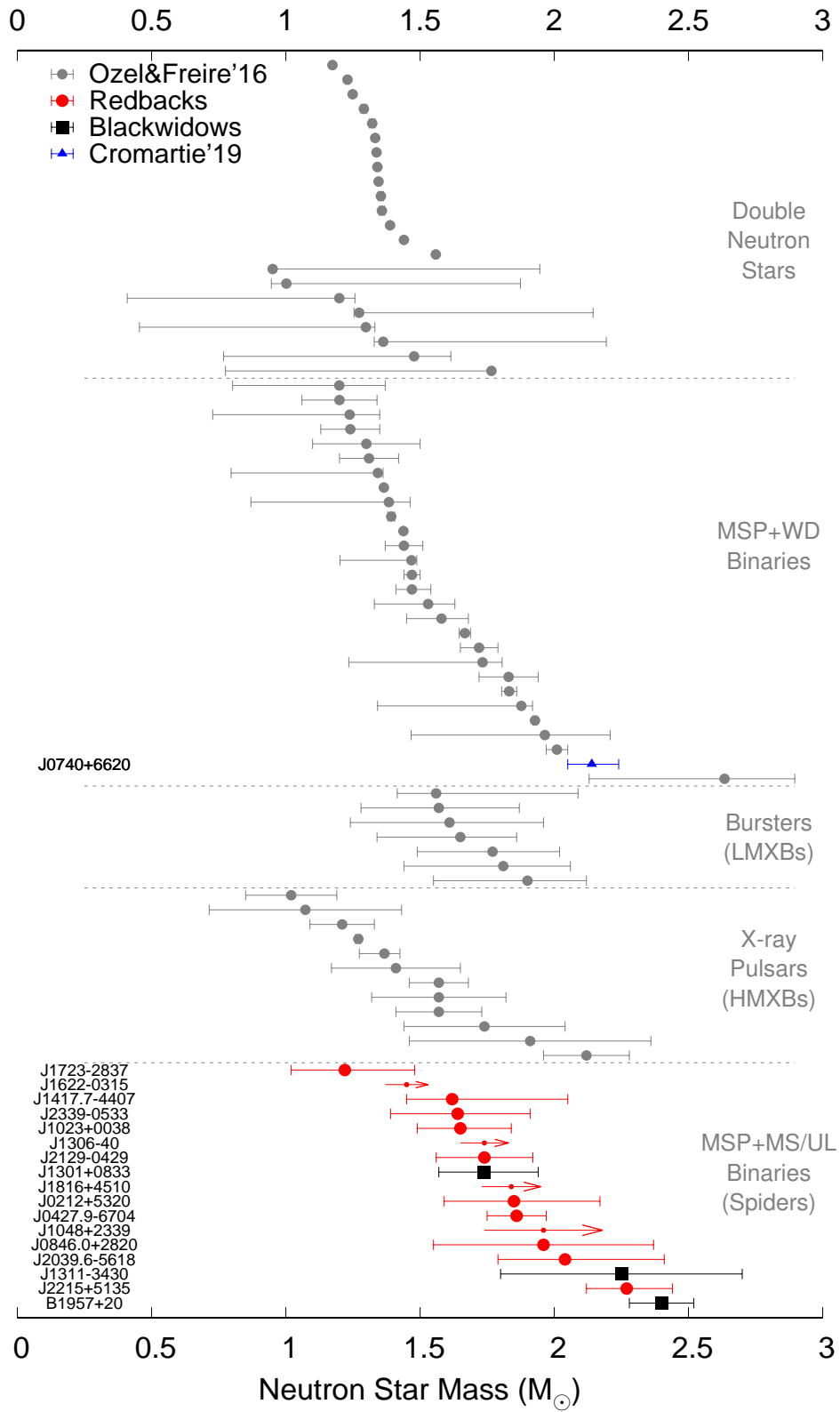}
    \vspace{-0.5cm}
\caption{Masses of 17 compact binary MSPs (red circles and black
  squares, as indicated; arrows show lower limits on M$_{NS}$;
  \citep{Romani16b,Shahbaz17,Romani15b,Linares18b,Kerkwijk11,Strader19}). For
  comparison, we show 68 NS mass measurements compiled by {\"O}zel \&
  Freire (\citep{Ozel16}, gray circles; see references
  therein). Different types of binary NSs are noted with labels along
  the right side. The recent NS mass measured by Cromartie et
  al. \citep{Cromartie19} is shown in blue.}
\label{fig:masses}
\end{center}
\end{figure}

\section{Super-massive neutron stars: breaking the 2~M$_\odot$ barrier}
\label{sec:massiveNS}

Figure~\ref{fig:masses} shows an updated compilation of 86 NS mass
measurements, including 17 compact binary MSPs
(\citep{Ozel16,Strader19,Cromartie19}, and references therein).
This readily shows that redback and black widow MSPs (or ``spiders'')
tend to push the NS mass range beyond the 2~M$_\odot$ limit.
In Table~\ref{table:masses} we collect those systems with NS mass
measurements above 2~M$_\odot$, giving the original references and
some details on each mass measurement.
In the following we discuss briefly these super-massive NSs, and argue
that compact binary MSPs are key to establishing what is the maximum
NS mass.

\begin{table}[ht]
\caption{Super-massive neutron stars, sorted in order of increasing
  mass measurement/constraint. Compact binary MSPs (spiders) are shown
  in boldface. Errors are at the 1-$\sigma$ confidence level (ranges
  indicate mass constraints rather than actual measurements). }
\begin{minipage}{\textwidth}
\centering
\begin{tabular}{$l ^c ^c ^c ^c ^c}  
\hline\hline
Name &
Type\footnote{RB: redback; RBc: redback candidate; BW: black widow; HMXB: high-mass X-ray binary; MSP+WD: millisecond pulsar with white dwarf companion (a question mark indicates that this WD identification is uncertain).} &
M$_\mathrm{NS}$ (M$_\odot$) &
Mean error &
P$_{orb}$ (hr) &
Ref. (M$_\mathrm{NS}$) \\
\hline
\hline
PSR~J0348+0432 & MSP+WD & 2.01$\pm$0.04 & 2\% & 2.5 & \citep{Antoniadis13}\\
  \rowstyle{\bfseries}
  3FGL~J2039.6-5618 & RBc & 2.04$^{{\bf+0.37}}_{{\bf-0.25}}$\footnote{Candidate RB, no pulsar detected yet. This M$_\mathrm{NS}$ comes from optical light curve modelling only, without considering radial velocities.} & 15\% & 5.5 & \citep{Strader19}\\
PSR~B1516+02B & MSP+WD? & 1.5--2.2\footnote{Note this is a probabilistic constraint assuming random inclination. We give latest reported range from {\"O}zel~\& Freire. Original 1-$\sigma$ range was 2.08$\pm$0.19~M$_\odot$.} & 19\% & 164.6 & \citep{Ozel16,Freire08b}\\
Vela~X-1 & HMXB & 2.12$\pm$0.16 & 8\% & 215.1 & \citep{Falanga15}\\
PSR J0740+6620 & MSP+WD & 2.14$^{+0.10}_{-0.09}$ & 4\% & 115.2 & \citep{Cromartie19}\\
  \rowstyle{\bfseries}
PSR~J1311--3430 & BW & 1.8--2.7\footnote{Formally 2.25$\pm$0.45~M$_\odot$, but authors give 1.8--2.7~M$_\odot$ range from a discussion of possible systematics, so this is a mass constraint (not a 1-$\sigma$ range).} & 20\%  & 1.6 & \citep{Romani15b}\\
  \rowstyle{\bfseries}
PSR~J2215+5135 & RB & 2.27$\pm$0.16 & 7\% & 4.1 & \citep{Linares18b}\\
  \rowstyle{\bfseries}
PSR~B1957+20 & BW & 2.40$\pm$0.12 & 5\% & 9.2 & \citep{Kerkwijk11}\\
PSR~J1748-2021B & MSP+WD? & 2.1--2.9\footnote{Note this is a probabilistic constraint assuming random inclination. We give latest reported range from {\"O}zel \& Freire. Original 1-$\sigma$ range was 2.74$\pm$0.21~M$_\odot$.} & 15\% & 493.2 & \citep{Ozel16,Freire08a}\\
\hline\hline
\end{tabular}
\end{minipage}
\label{table:masses}
\end{table}

As mentioned in Section~\ref{sec:intro}, the first-discovered compact
binary MSP \citep{Fruchter88}, the original black widow pulsar
PSR~B1957+20, holds the current record of the highest NS mass
measurement, with M$_\mathrm{NS}$=2.40$\pm$0.12 M$_\odot$
\citep{Kerkwijk11}.
As discussed by van Kerkwijk et al., this result may be sensitive to
systematic uncertainties on the orbital inclination (i=65$^\circ\pm$2,
which the authors take from earlier light curve modelling results;
\citep{Reynolds07}) and on the center of mass velocity (estimated from
a combination of simulated spectra and analytical arguments
(\citep{Kerkwijk11}; see their Section~5).
An independent analysis and joint modeling of the light and radial
velocity curves, in light of recent results, should be able to
establish or update this mass measurement.
In their final analysis of the light and radial velocity curves of the
black widow PSR~J1311-3430, Romani et al. report a NS mass in the
range 1.8--2.7~M$_\odot$ and argue that the accuracy of this
measurement is partly limited by optical flares and variable wind
emission lines \citep{Romani15b}.

The mass of the X-ray pulsar in the high-mass X-ray binary (HMXB)
Vela~X-1 has also been found to be higher than 2~M$_\odot$:
2.12$\pm$0.16~M$_\odot$ \citep{Falanga15}.
Even though the optical determination of the companion's velocity may
suffer from additional systematics, most measurements indicate that
this HMXB hosts a massive NS \citep{Kerkwijk95,Quaintrell03}.
In globular clusters, dynamical measurements of the companion star are
challenging, but there is also evidence for super-massive NSs
\citep{Freire08a,Freire08b}.
Precise long-term timing of radio MSPs in wide eccentric binaries has
allowed highly significant detections of the rate of advance of
periastron.
If this advance is exclusively due to general-relativistic periastron
precession, such detections lead to an accurate measurement of the
total mass in the binary \citep{Ozel12}.
Even if the companion is undetected and the orbital inclination is
unknown, Freire et al. used probabilistic arguments to constrain the
masses of two such pulsars to be close to or above 2~M$_\odot$
(PSR~J1748-2021B and PSR~B1516+02B, see Table~\ref{table:masses}).




Other redback MSPs have been found that may be close to or above the
2~M$_\odot$ limit.
Strader and collaborators set a rather high lower limit on the mass of
PSR J1048+2339: M$_\mathrm{NS}$$>$1.96$\pm$0.22~M$_\odot$
(\citep{Strader19}; as they note, the relatively large uncertainty on
this limit can be improved with more precise dynamical studies).
The redback candidate 3FGL~J2039.6-5618 constitutes a similar case: a
measurement close to 2~M$_\odot$ but with a relatively large
statistical uncertainty (2.04$^{+0.37}_{-0.25}$~M$_\odot$).
Because pulsations have not been detected and this measurement is based
solely on light curve modeling (without radial velocity measurements
available so far), the implications for M$_{NS}^{max}$ are less
certain.


%
As mentioned in Section~\ref{sec:intro}, Antoniadis and collaborators
found a NS mass of 2.01$\pm$0.04~M$_\odot$ using pulsar timing and
optical observations of PSR~J0348+0432 (modeling the WD absorption
lines to find M$_c$; see \citep{Antoniadis13}).
Most recently, Cromartie et al. \citep{Cromartie19} found in
PSR~J0740+6620 another super-massive NS with a mass above 2~M$_\odot$:
M$_\mathrm{NS}$=2.14$\pm$0.10~M$_\odot$.
This result comes from a different method to measure M$_\mathrm{NS}$,
i.e., the measurement of the relativistic Shapiro delay (see also
\citep{Demorest10,Fonseca16}).
Finding super-massive NSs with independent measurement techniques is
important to establish their maximum mass, and to rule out systematic
effects in the mass determination (ideally, applying these different
techniques to the same pulsar).

\begin{figure}[h!]
  \begin{center}
    \resizebox{0.85\columnwidth}{!}{\rotatebox{-90}{\includegraphics[]{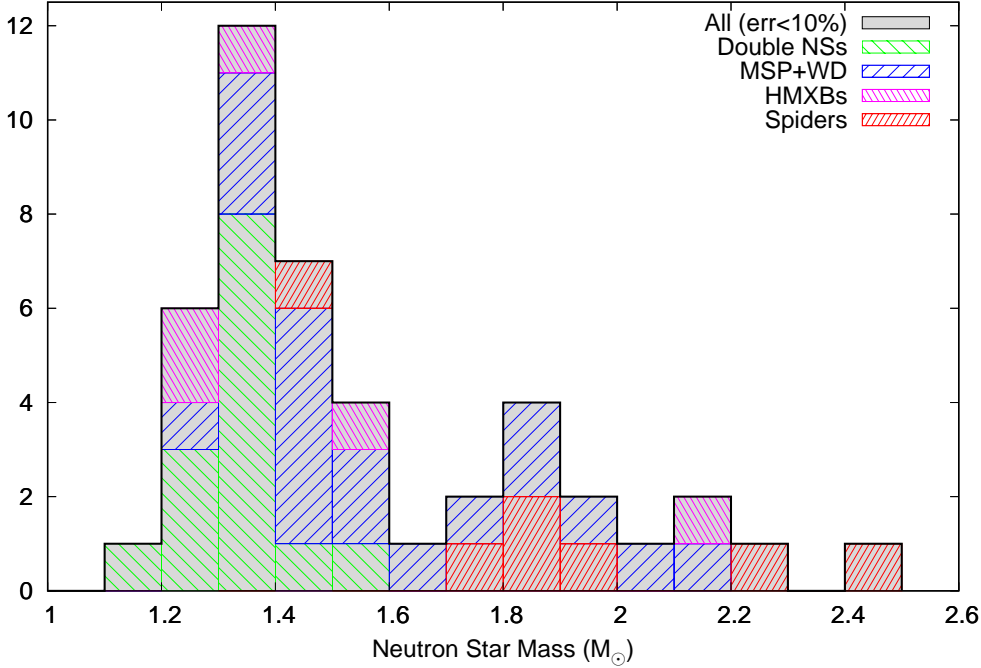}}}
\caption{Histogram of the 44 NS mass measurements with errors smaller
  than 10\%, showing the different source classes. While double NSs
  cluster around 1.35 M$_\odot$, spiders have masses from 1.4 to 2.4
  M$_\odot$.}
\label{fig:MNShist}
\end{center}
\end{figure}

\section{Conclusions}
\label{sec:conclusions}

To summarize, we show in Figure~\ref{fig:MNShist} a histogram of NS
masses, using only mass measurements with fractional uncertainties
below 10\% (which includes 44 out of the 86 systems presented in
Figure~\ref{fig:masses}).
It has been already pointed out that double NSs show a narrow mass
distribution centered around 1.35~M$_\odot$ and that recycled
fast-spinning MSPs are on average more massive
\citep{Thorsett99,Lattimer07,Ozel12,Ozel16}.
During the past decade, we have found mounting evidence for
super-massive NSs with masses in the range 2--2.5~M$_\odot$
(Table~\ref{table:masses}).
Here we stress that the growing population of compact binary MSPs
shows great potential for finding super-massive NSs and therefore
constraining M$_{NS}^{max}$ and the EoS of ultradense matter.
Indeed, 4 of the 9 systems with potentially super-massive NSs shown in
Table~\ref{table:masses} are redbacks (one of them candidate) or black
widows, and this number may well increase as the population of compact
binary MSPs continues to grow.
Newly discovered nearby redback MSPs, in particular, hold the greatest
potential for the next decade, as they have moderately low-mass
companion stars which are sufficiently bright in the optical band
(typically $\sim$18-20 mag) to pursue high S/N spectroscopy and
detailed dynamical studies.


\bibliography{../../biblio.bib}

\providecommand{\href}[2]{#2}\begingroup\raggedright\begin{thebibliography}{10}

\bibitem{Casares94}
J.~{Casares} and P.~A. {Charles}, \emph{Optical studies of V404 Cyg, the X-ray
  transient GS 2023+338. IV. The rotation speed of the companion star.},
  {\emph{\mnras} {\bfseries 271} (1994) L5}.

\bibitem{Timmes96}
F.~X. {Timmes}, S.~E. {Woosley} and T.~A. {Weaver}, \emph{The Neutron Star and
  Black Hole Initial Mass Function},
  \href{https://doi.org/10.1086/176778}{\emph{\apj} {\bfseries 457} (1996) 834}
  [\href{https://arxiv.org/abs/astro-ph/9510136}{{\ttfamily
  astro-ph/9510136}}].

\bibitem{Alpar82}
M.~A. {Alpar}, A.~F. {Cheng}, M.~A. {Ruderman} and J.~{Shaham}, \emph{A new
  class of radio pulsars}, \href{https://doi.org/10.1038/300728a0}{\emph{\nat}
  {\bfseries 300} (1982) 728}.

\bibitem{GW170817}
{\scshape LIGO Scientific Collaboration and Virgo Collaboration} collaboration,
  \emph{GW170817: Observation of Gravitational Waves from a Binary Neutron Star
  Inspiral}, \href{https://doi.org/10.1103/PhysRevLett.119.161101}{\emph{Phys.
  Rev. Lett.} {\bfseries 119} (2017) 161101}.

\bibitem{Margalit17}
B.~{Margalit} and B.~D. {Metzger}, \emph{Constraining the Maximum Mass of
  Neutron Stars from Multi-messenger Observations of GW170817},
  \href{https://doi.org/10.3847/2041-8213/aa991c}{\emph{\apjl} {\bfseries 850}
  (2017) L19} [\href{https://arxiv.org/abs/1710.05938}{{\ttfamily
  1710.05938}}].

\bibitem{Salpeter60}
E.~Salpeter, \emph{Matter at high densities},
  \href{https://doi.org/https://doi.org/10.1016/0003-4916(60)90006-3}{\emph{Annals
  of Physics} {\bfseries 11} (1960) 393 }.

\bibitem{Page06}
D.~{Page} and S.~{Reddy}, \emph{Dense Matter in Compact Stars: Theoretical
  Developments and Observational Constraints},
  \href{https://doi.org/10.1146/annurev.nucl.56.080805.140600}{\emph{Annual
  Review of Nuclear and Particle Science} {\bfseries 56} (2006) 327}
  [\href{https://arxiv.org/abs/astro-ph/0608360}{{\ttfamily
  astro-ph/0608360}}].

\bibitem{Lattimer07}
J.~M. {Lattimer} and M.~{Prakash}, \emph{Neutron star observations: Prognosis
  for equation of state constraints},
  \href{https://doi.org/10.1016/j.physrep.2007.02.003}{\emph{\physrep}
  {\bfseries 442} (2007) 109}
  [\href{https://arxiv.org/abs/arXiv:astro-ph/0612440}{{\ttfamily
  arXiv:astro-ph/0612440}}].

\bibitem{Ozel16}
F.~{{\"O}zel} and P.~{Freire}, \emph{Masses, Radii, and the Equation of State
  of Neutron Stars},
  \href{https://doi.org/10.1146/annurev-astro-081915-023322}{\emph{\araa}
  {\bfseries 54} (2016) 401}
  [\href{https://arxiv.org/abs/1603.02698}{{\ttfamily 1603.02698}}].

\bibitem{ATNFpsrcat}
R.~N. {Manchester}, G.~B. {Hobbs}, A.~{Teoh} and M.~{Hobbs}, \emph{The
  Australia Telescope National Facility Pulsar Catalogue},
  \href{https://doi.org/10.1086/428488}{\emph{\aj} {\bfseries 129} (2005) 1993}
  [\href{https://arxiv.org/abs/astro-ph/0412641}{{\ttfamily
  astro-ph/0412641}}].

\bibitem{LorimerCat}
D.~{Lorimer}, \emph{Galactic Millisecond Pulsars}, {\emph{Online catalog at
  http://astro.phys.wvu.edu/GalacticMSPs} (2019) }.

\bibitem{Thorsett99}
S.~E. {Thorsett} and D.~{Chakrabarty}, \emph{Neutron Star Mass Measurements. I.
  Radio Pulsars}, \href{https://doi.org/10.1086/306742}{\emph{\apj} {\bfseries
  512} (1999) 288}
  [\href{https://arxiv.org/abs/arXiv:astro-ph/9803260}{{\ttfamily
  arXiv:astro-ph/9803260}}].

\bibitem{Demorest10}
P.~B. {Demorest}, T.~{Pennucci}, S.~M. {Ransom}, M.~S.~E. {Roberts} and
  J.~W.~T. {Hessels}, \emph{A two-solar-mass neutron star measured using
  Shapiro delay}, \href{https://doi.org/10.1038/nature09466}{\emph{\nat}
  {\bfseries 467} (2010) 1081}
  [\href{https://arxiv.org/abs/1010.5788}{{\ttfamily 1010.5788}}].

\bibitem{Antoniadis13}
J.~{Antoniadis}, P.~C.~C. {Freire}, N.~{Wex}, T.~M. {Tauris}, R.~S. {Lynch},
  M.~H. {van Kerkwijk} et~al., \emph{A Massive Pulsar in a Compact Relativistic
  Binary}, \href{https://doi.org/10.1126/science.1233232}{\emph{Science}
  {\bfseries 340} (2013) 448}
  [\href{https://arxiv.org/abs/1304.6875}{{\ttfamily 1304.6875}}].

\bibitem{Kerkwijk11}
M.~H. {van Kerkwijk}, R.~P. {Breton} and S.~R. {Kulkarni}, \emph{Evidence for a
  Massive Neutron Star from a Radial-velocity Study of the Companion to the
  Black-widow Pulsar PSR B1957+20},
  \href{https://doi.org/10.1088/0004-637X/728/2/95}{\emph{\apj} {\bfseries 728}
  (2011) 95} [\href{https://arxiv.org/abs/1009.5427}{{\ttfamily 1009.5427}}].

\bibitem{Ozel12}
F.~{{\"O}zel}, D.~{Psaltis}, R.~{Narayan} and A.~{Santos Villarreal}, \emph{On
  the Mass Distribution and Birth Masses of Neutron Stars},
  \href{https://doi.org/10.1088/0004-637X/757/1/55}{\emph{\apj} {\bfseries 757}
  (2012) 55} [\href{https://arxiv.org/abs/1201.1006}{{\ttfamily 1201.1006}}].

\bibitem{Romani12b}
R.~W. {Romani}, A.~V. {Filippenko}, J.~M. {Silverman}, S.~B. {Cenko},
  J.~{Greiner}, A.~{Rau} et~al., \emph{PSR J1311-3430: A Heavyweight Neutron
  Star with a Flyweight Helium Companion},
  \href{https://doi.org/10.1088/2041-8205/760/2/L36}{\emph{\apjl} {\bfseries
  760} (2012) L36} [\href{https://arxiv.org/abs/1210.6884}{{\ttfamily
  1210.6884}}].

\bibitem{Breton13}
R.~P. {Breton}, M.~H. {van Kerkwijk}, M.~S.~E. {Roberts}, J.~W.~T. {Hessels},
  F.~{Camilo}, M.~A. {McLaughlin} et~al., \emph{Discovery of the Optical
  Counterparts to Four Energetic Fermi Millisecond Pulsars},
  \href{https://doi.org/10.1088/0004-637X/769/2/108}{\emph{\apj} {\bfseries
  769} (2013) 108} [\href{https://arxiv.org/abs/1302.1790}{{\ttfamily
  1302.1790}}].

\bibitem{Linares18b}
M.~{Linares}, T.~{Shahbaz} and J.~{Casares}, \emph{Peering into the Dark Side:
  Magnesium Lines Establish a Massive Neutron Star in PSR J2215+5135},
  \href{https://doi.org/10.3847/1538-4357/aabde6}{\emph{\apj} {\bfseries 859}
  (2018) 54} [\href{https://arxiv.org/abs/1805.08799}{{\ttfamily 1805.08799}}].

\bibitem{Strader19}
J.~{Strader}, S.~{Swihart}, L.~{Chomiuk}, A.~{Bahramian}, C.~{Britt}, C.~C.
  {Cheung} et~al., \emph{Optical Spectroscopy and Demographics of Redback
  Millisecond Pulsar Binaries},
  \href{https://doi.org/10.3847/1538-4357/aafbaa}{\emph{\apj} {\bfseries 872}
  (2019) 42} [\href{https://arxiv.org/abs/1812.04626}{{\ttfamily 1812.04626}}].

\bibitem{FreireCat}
P.~{Freire}, \emph{Pulsars in Globular Clusters}, {\emph{Online catalog at
  http://www.naic.edu/pfreire/GCpsr.html} (2019) }.

\bibitem{Acero15}
F.~{Acero et al.}, \emph{Fermi Large Area Telescope Third Source Catalog},
  \href{https://doi.org/10.1088/0067-0049/218/2/23}{\emph{\apjs} {\bfseries
  218} (2015) 23} [\href{https://arxiv.org/abs/1501.02003}{{\ttfamily
  1501.02003}}].

\bibitem{4FGL}
{The Fermi-LAT collaboration}, \emph{Fermi Large Area Telescope Fourth Source
  Catalog}, {\emph{arXiv e-prints} (2019) arXiv:1902.10045}
  [\href{https://arxiv.org/abs/1902.10045}{{\ttfamily 1902.10045}}].

\bibitem{Hessels11}
J.~W.~T. {Hessels}, M.~S.~E. {Roberts}, M.~A. {McLaughlin}, P.~S. {Ray},
  P.~{Bangale}, S.~M. {Ransom} et~al., \emph{A 350-MHz GBT Survey of 50 Faint
  Fermi {$\gamma$}-ray Sources for Radio Millisecond Pulsars},  in
  \emph{American Institute of Physics Conference Series}, M.~{Burgay},
  N.~{D'Amico}, P.~{Esposito}, A.~{Pellizzoni} and A.~{Possenti}, eds.,
  vol.~1357 of \emph{American Institute of Physics Conference Series},
  pp.~40--43, 2011, \href{https://arxiv.org/abs/1101.1742}{{\ttfamily
  1101.1742}}, \href{https://doi.org/10.1063/1.3615072}{DOI}.

\bibitem{Ransom11}
S.~M. {Ransom}, P.~S. {Ray}, F.~{Camilo}, M.~S.~E. {Roberts},
  {\"O}.~{{\c{C}}elik}, M.~T. {Wolff} et~al., \emph{Three Millisecond Pulsars
  in Fermi LAT Unassociated Bright Sources},
  \href{https://doi.org/10.1088/2041-8205/727/1/L16}{\emph{\apjl} {\bfseries
  727} (2011) L16} [\href{https://arxiv.org/abs/1012.2862}{{\ttfamily
  1012.2862}}].

\bibitem{Bates11}
S.~D. {Bates}, M.~{Bailes}, N.~D.~R. {Bhat}, M.~{Burgay}, S.~{Burke-Spolaor},
  N.~{D'Amico} et~al., \emph{The High Time Resolution Universe Pulsar Survey -
  II. Discovery of five millisecond pulsars},
  \href{https://doi.org/10.1111/j.1365-2966.2011.18416.x}{\emph{\mnras}
  {\bfseries 416} (2011) 2455}
  [\href{https://arxiv.org/abs/1101.4778}{{\ttfamily 1101.4778}}].

\bibitem{Ray12}
P.~S. {Ray}, A.~A. {Abdo}, D.~{Parent}, D.~{Bhattacharya}, B.~{Bhattacharyya},
  F.~{Camilo} et~al., \emph{Radio Searches of Fermi LAT Sources and Blind
  Search Pulsars: The Fermi Pulsar Search Consortium}, {\emph{2011 Fermi
  Symposium proceedings - eConf C110509; ArXiv 1205.3089} (2012) }
  [\href{https://arxiv.org/abs/1205.3089}{{\ttfamily 1205.3089}}].

\bibitem{Stovall14}
K.~{Stovall}, R.~S. {Lynch}, S.~M. {Ransom}, A.~M. {Archibald}, S.~{Banaszak},
  C.~M. {Biwer} et~al., \emph{The Green Bank Northern Celestial Cap Pulsar
  Survey. I. Survey Description, Data Analysis, and Initial Results},
  \href{https://doi.org/10.1088/0004-637X/791/1/67}{\emph{\apj} {\bfseries 791}
  (2014) 67} [\href{https://arxiv.org/abs/1406.5214}{{\ttfamily 1406.5214}}].

\bibitem{Camilo15}
F.~{Camilo}, M.~{Kerr}, P.~S. {Ray}, S.~M. {Ransom}, J.~{Sarkissian}, H.~T.
  {Cromartie} et~al., \emph{Parkes Radio Searches of Fermi Gamma-Ray Sources
  and Millisecond Pulsar Discoveries},
  \href{https://doi.org/10.1088/0004-637X/810/2/85}{\emph{The Astrophysical
  Journal} {\bfseries 810} (2015) 85}
  [\href{https://arxiv.org/abs/1507.04451}{{\ttfamily 1507.04451}}].

\bibitem{Bates15}
S.~D. {Bates}, D.~{Thornton}, M.~{Bailes}, E.~{Barr}, C.~G. {Bassa}, N.~D.~R.
  {Bhat} et~al., \emph{The High Time Resolution Universe survey - XI. Discovery
  of five recycled pulsars and the optical detectability of survey white dwarf
  companions}, \href{https://doi.org/10.1093/mnras/stu2350}{\emph{\mnras}
  {\bfseries 446} (2015) 4019}
  [\href{https://arxiv.org/abs/1411.1288}{{\ttfamily 1411.1288}}].

\bibitem{Sanpa16}
S.~{Sanpa-Arsa}, \emph{Searching for New Millisecond Pulsars with the GBT in
  Fermi Unassociated Sources}, {\emph{PhD Thesis, University of Virginia}
  (2016) }.

\bibitem{Romani11}
R.~W. {Romani} and M.~S. {Shaw}, \emph{The Orbit and Companion of Probable
  {$\gamma$}-Ray Pulsar J2339-0533},
  \href{https://doi.org/10.1088/2041-8205/743/2/L26}{\emph{\apjl} {\bfseries
  743} (2011) L26} [\href{https://arxiv.org/abs/1111.3074}{{\ttfamily
  1111.3074}}].

\bibitem{Kong12}
A.~K.~H. {Kong}, R.~H.~H. {Huang}, K.~S. {Cheng}, J.~{Takata}, Y.~{Yatsu},
  C.~C. {Cheung} et~al., \emph{Discovery of an Unidentified Fermi Object as a
  Black Widow-like Millisecond Pulsar},
  \href{https://doi.org/10.1088/2041-8205/747/1/L3}{\emph{\apjl} {\bfseries
  747} (2012) L3} [\href{https://arxiv.org/abs/1201.3629}{{\ttfamily
  1201.3629}}].

\bibitem{Romani12}
R.~W. {Romani}, \emph{2FGL J1311.7-3429 Joins the Black Widow Club},
  \href{https://doi.org/10.1088/2041-8205/754/2/L25}{\emph{\apjl} {\bfseries
  754} (2012) L25} [\href{https://arxiv.org/abs/1207.1736}{{\ttfamily
  1207.1736}}].

\bibitem{Romani14}
R.~W. {Romani}, A.~V. {Filippenko} and S.~B. {Cenko}, \emph{2FGL J1653.6-0159:
  A New Low in Evaporating Pulsar Binary Periods},
  \href{https://doi.org/10.1088/2041-8205/793/1/L20}{\emph{\apjl} {\bfseries
  793} (2014) L20} [\href{https://arxiv.org/abs/1408.2886}{{\ttfamily
  1408.2886}}].

\bibitem{Li16}
K.-L. {Li}, A.~K.~H. {Kong}, X.~{Hou}, J.~{Mao}, J.~{Strader}, L.~{Chomiuk}
  et~al., \emph{Discovery of a Redback Millisecond Pulsar Candidate: 3FGL
  J0212.1+5320}, \href{https://doi.org/10.3847/1538-4357/833/2/143}{\emph{\apj}
  {\bfseries 833} (2016) 143}
  [\href{https://arxiv.org/abs/1609.02951}{{\ttfamily 1609.02951}}].

\bibitem{Linares17}
M.~{Linares}, P.~{Miles-P{\'a}ez}, P.~{Rodr{\'{\i}}guez-Gil}, T.~{Shahbaz},
  J.~{Casares}, C.~{Fari{\~n}a} et~al., \emph{A millisecond pulsar candidate in
  a 21-h orbit: 3FGL J0212.1+5320},
  \href{https://doi.org/10.1093/mnras/stw3057}{\emph{\mnras} {\bfseries 465}
  (2017) 4602} [\href{https://arxiv.org/abs/1609.02232}{{\ttfamily
  1609.02232}}].

\bibitem{Chen13}
H.-L. {Chen}, X.~{Chen}, T.~M. {Tauris} and Z.~{Han}, \emph{Formation of Black
  Widows and Redbacks -- Two Distinct Populations of Eclipsing Binary
  Millisecond Pulsars},
  \href{https://doi.org/10.1088/0004-637X/775/1/27}{\emph{\apj} {\bfseries 775}
  (2013) 27} [\href{https://arxiv.org/abs/1308.4107}{{\ttfamily 1308.4107}}].

\bibitem{Schroeder14}
J.~{Schroeder} and J.~{Halpern}, \emph{Observations and Modeling of the
  Companions of Short Period Binary Millisecond Pulsars: Evidence for High-mass
  Neutron Stars}, \href{https://doi.org/10.1088/0004-637X/793/2/78}{\emph{\apj}
  {\bfseries 793} (2014) 78} [\href{https://arxiv.org/abs/1401.7966}{{\ttfamily
  1401.7966}}].

\bibitem{Romani15}
R.~W. {Romani}, M.~L. {Graham}, A.~V. {Filippenko} and M.~{Kerr}, \emph{Keck
  Spectroscopy of Millisecond Pulsar J2215+5135: A Moderate-M$_{NS}$,
  High-inclination Binary},
  \href{https://doi.org/10.1088/2041-8205/809/1/L10}{\emph{\apjl} {\bfseries
  809} (2015) L10} [\href{https://arxiv.org/abs/1506.04332}{{\ttfamily
  1506.04332}}].

\bibitem{Romani16}
R.~W. {Romani} and N.~{Sanchez}, \emph{Intra-binary Shock Heating of Black
  Widow Companions},
  \href{https://doi.org/10.3847/0004-637X/828/1/7}{\emph{\apj} {\bfseries 828}
  (2016) 7} [\href{https://arxiv.org/abs/1606.03518}{{\ttfamily 1606.03518}}].

\bibitem{Romani16b}
R.~W. {Romani}, M.~L. {Graham}, A.~V. {Filippenko} and W.~{Zheng}, \emph{PSR
  J1301+0833: A Kinematic Study of a Black-widow Pulsar},
  \href{https://doi.org/10.3847/1538-4357/833/2/138}{\emph{\apj} {\bfseries
  833} (2016) 138}.

\bibitem{Shahbaz17}
T.~{Shahbaz}, M.~{Linares} and R.~P. {Breton}, \emph{Properties of the redback
  millisecond pulsar binary 3FGL J0212.1+5320},
  \href{https://doi.org/10.1093/mnras/stx2195}{\emph{\mnras} {\bfseries 472}
  (2017) 4287} [\href{https://arxiv.org/abs/1708.07355}{{\ttfamily
  1708.07355}}].

\bibitem{Romani15b}
R.~W. {Romani}, A.~V. {Filippenko} and S.~B. {Cenko}, \emph{A Spectroscopic
  Study of the Extreme Black Widow PSR J1311-3430},
  \href{https://doi.org/10.1088/0004-637X/804/2/115}{\emph{\apj} {\bfseries
  804} (2015) 115} [\href{https://arxiv.org/abs/1503.05247}{{\ttfamily
  1503.05247}}].

\bibitem{Cromartie19}
H.~T. {Cromartie}, E.~{Fonseca}, S.~M. {Ransom}, P.~B. {Demorest},
  Z.~{Arzoumanian}, H.~{Blumer} et~al., \emph{Relativistic Shapiro delay
  measurements of an extremely massive millisecond pulsar}, {\emph{Nature
  Astronomy} (2019) arXiv:1904.06759}
  [\href{https://arxiv.org/abs/1904.06759}{{\ttfamily 1904.06759}}].

\bibitem{Freire08b}
P.~C.~C. {Freire}, A.~{Wolszczan}, M.~{van den Berg} and J.~W.~T. {Hessels},
  \emph{A Massive Neutron Star in the Globular Cluster M5},
  \href{https://doi.org/10.1086/587832}{\emph{\apj} {\bfseries 679} (2008)
  1433} [\href{https://arxiv.org/abs/0712.3826}{{\ttfamily 0712.3826}}].

\bibitem{Falanga15}
M.~{Falanga}, E.~{Bozzo}, A.~{Lutovinov}, J.~M. {Bonnet-Bidaud}, Y.~{Fetisova}
  and J.~{Puls}, \emph{Ephemeris, orbital decay, and masses of ten eclipsing
  high-mass X-ray binaries},
  \href{https://doi.org/10.1051/0004-6361/201425191}{\emph{\aap} {\bfseries
  577} (2015) A130} [\href{https://arxiv.org/abs/1502.07126}{{\ttfamily
  1502.07126}}].

\bibitem{Freire08a}
P.~C.~C. {Freire}, S.~M. {Ransom}, S.~{B{\'e}gin}, I.~H. {Stairs}, J.~W.~T.
  {Hessels}, L.~H. {Frey} et~al., \emph{Eight New Millisecond Pulsars in NGC
  6440 and NGC 6441}, \href{https://doi.org/10.1086/526338}{\emph{\apj}
  {\bfseries 675} (2008) 670}
  [\href{https://arxiv.org/abs/0711.0925}{{\ttfamily 0711.0925}}].

\bibitem{Fruchter88}
A.~S. {Fruchter}, D.~R. {Stinebring} and J.~H. {Taylor}, \emph{A millisecond
  pulsar in an eclipsing binary},
  \href{https://doi.org/10.1038/333237a0}{\emph{\nat} {\bfseries 333} (1988)
  237}.

\bibitem{Reynolds07}
M.~T. {Reynolds}, P.~J. {Callanan}, A.~S. {Fruchter}, M.~A.~P. {Torres}, M.~E.
  {Beer} and R.~A. {Gibbons}, \emph{The light curve of the companion to PSR
  B1957+20},
  \href{https://doi.org/10.1111/j.1365-2966.2007.11991.x}{\emph{\mnras}
  {\bfseries 379} (2007) 1117}
  [\href{https://arxiv.org/abs/0705.2514}{{\ttfamily 0705.2514}}].

\bibitem{Kerkwijk95}
M.~H. {van Kerkwijk}, J.~{van Paradijs} and E.~J. {Zuiderwijk}, \emph{On the
  masses of neutron stars.}, {\emph{\aap} {\bfseries 303} (1995) 497}
  [\href{https://arxiv.org/abs/astro-ph/9505071}{{\ttfamily
  astro-ph/9505071}}].

\bibitem{Quaintrell03}
H.~{Quaintrell}, A.~J. {Norton}, T.~D.~C. {Ash}, P.~{Roche}, B.~{Willems},
  T.~R. {Bedding} et~al., \emph{The mass of the neutron star in Vela X-1 and
  tidally induced non-radial oscillations in GP Vel},
  \href{https://doi.org/10.1051/0004-6361:20030120}{\emph{\aap} {\bfseries 401}
  (2003) 313} [\href{https://arxiv.org/abs/arXiv:astro-ph/0301243}{{\ttfamily
  arXiv:astro-ph/0301243}}].

\bibitem{Fonseca16}
E.~{Fonseca}, T.~T. {Pennucci}, J.~A. {Ellis}, I.~H. {Stairs}, D.~J. {Nice},
  S.~M. {Ransom} et~al., \emph{The NANOGrav Nine-year Data Set: Mass and
  Geometric Measurements of Binary Millisecond Pulsars},
  \href{https://doi.org/10.3847/0004-637X/832/2/167}{\emph{\apj} {\bfseries
  832} (2016) 167} [\href{https://arxiv.org/abs/1603.00545}{{\ttfamily
  1603.00545}}].

\end{thebibliography}\endgroup

\bigskip
\bigskip
\noindent {\bf DISCUSSION}

\bigskip
\noindent {\bf JOSEP MARIA PAREDES:} Are the orbital period and the
companion star's rotational period synchronized?

\bigskip
\noindent {\bf MANUEL LINARES:} Yes, as a result of tidal forces the
orbits are circular and the companion is ``tidally locked''.

\bigskip
\noindent {\bf SANDRO MEREGHETTI:} a) Is there any correlation between
pulsar mass and other system parameters? b) In the case of
PSR~J2215+5135 (and super-massive NSs in general): how much of that
mass has been accreted?

\bigskip
\noindent {\bf MANUEL LINARES:} a) Not that we know of. b) According
to evolutionary models (e.g. Chen et al. 2013), redbacks can accrete
up to 0.3--0.5~M$_\odot$ during their active accretion phase. This
suggests that the pulsars would have to be born massive to go beyond
2~$M_\odot$.

\bigskip
\noindent {\bf ANDREA SANTANGELO:} Some quark-based equations of state
for the NS core do allow for super-massive NSs with
M$_{NS}$>2~M$_\odot$.

\bigskip
\noindent {\bf MANUEL LINARES:} Yes, hyperons seem ruled out but
deconfined quarks may be able to do it (especially in the most recent
quark EoSs).


\end{document}